# One-Sided Device-Independent Random Number Generation Through Fiber Channels

Jinfang Zhang,[1,][*] Yi Li,[2, 3,][*] Mengyu Zhao,[1] Dongmei Han,[1] Jun Liu,[1] Meihong Wang,[1,4] Qihuang Gong,[2,4,5] Yu Xiang,[2,4,][†] Qiongyi He,[2,4,5,6] and Xiaolong Su[1,4,][‡]

[1] State Key Laboratory of Quantum Optics and Quantum Optics Devices,
Institute of Opto-Electronics, Shanxi University, Taiyuan, 030006, China

[2] State Key Laboratory for Mesoscopic Physics, School of Physics, Frontiers Science Center for Nano-optoelectronics,
& Collaborative Innovation Center of Quantum Matter, Peking University, Beijing 100871, China

[3] Beijing Academy of Quantum Information Sciences, Beijing 100193, China

[4] Collaborative Innovation Center of Extreme Optics,
Shanxi University, Taiyuan, Shanxi 030006, China

[5] Peking University Yangtze Delta Institute of Optoelectronics, Nantong, Jiangsu 226010, China

[6] Hefei National Laboratory, Hefei 230088, China

Randomness is an essential resource and plays important roles in various applications ranging from cryptography to simulation of complex systems. Certified randomness from quantum process is ensured to have the element of privacy but usually relies on the device's behavior. To certify randomness without the characterization for device, it is crucial to realize the one-sided device-independent random number generation based on quantum steering, which guarantees security of randomness and relaxes the demands of one party's device. Here, we distribute quantum steering between two distant users through a 2 km fiber channel and generate quantum random numbers at the remote station with untrustworthy device. We certify the steering-based randomness by reconstructing covariance matrix of the Gaussian entangled state shared between distant parties. Then, the quantum random numbers with a generation rate of 7.06 Mbits/s are extracted from the measured amplitude quadrature fluctuation of the state owned by the remote party. Our results demonstrate the first realization of steering-based random numbers extraction in a practical fiber channel, which paves the way to the quantum random numbers generation in asymmetric networks.

**Introduction**

The random number has broad applications in many fields, ranging from simulation, cryptography, as well as computer networks or lotteries [1,2]. The randomness and its privacy require assumptions about the computational power of the adversary, which motivates the extensive study of true random number generators based on some unpredictable physical processes[3]. As a typical case of true random numbers, quantum random numbers have attracted much attention since the inherent randomness of measurement outputs in quantum mechanics is guaranteed by Born's rule[4]. For instance, random numbers can be generated by detecting single photons after they pass through a balanced beam-splitter[5], which relies on the device working in a particular way. However, deviations in the device behavior can affect the randomness of outputs and are difficult to detect, and any real device is too complicated to model in its entirety, which will leave open the possibility that an adversary can exploit a feature of the device outside the model, leading to the insecurity of generated random numbers[6].

This problem can be tackled by verifying Bell nonlocality[7] or Einstein-Podolsky-Rosen (EPR) steering[8] between distant users. The violation of Bell inequality ensures the quantum random number generation (QRNG) in a device-independent manner[9-13], where the user's devices are all untrusted. Different from Bell nonlocality, EPR steering can be used to realize one-sided device-independent (1SDI) QRNG, where only the device of steering party is untrusted since it is an asymmetric form of correlation that lies between entanglement[14] and Bell nonlocality[15] hierarchically. Such a 1SDI scenario typically allows users to own asymmetric devices, for instance, the bank and its clients where the bank owns trustworthy devices but the clients only have cheap and untrustworthy devices[16]. Furthermore, in a practical scenario, not all users involved in the network might have the ability to prepare the required quantum resources, especially for some remote untrusted users. In this case, entanglement resources can be prepared in a quantum server and distributed to remote users. Considering the practical channel loss and noise, how to verify the security and efficiency of the generated random numbers based on the shared quantum correlations remains a challenge.

Several experiments have implemented the randomness certification by allowing source or measurement devices to be partially characterized in the discrete variable system[17-22]. The deterministic preparation of continuous variable (CV) system benefits quantum random number generation[23]. For example, some source-device-independent protocols in CV systems can reaching a higher generation rate while removing the assumption about the source for bounding the side information[24-26]. Recently, a protocol of certifying the steering-based randomness in a CV system is demonstrated in a proof-of-principle experiment[27]. Towards practical applications, it is essential to certify randomness and extract quantum random numbers in a remote station based on the shared EPR steering.

In this work, we demonstrate the distribution of CV EPR steering over a 2 km fiber channel and extract quantum random numbers after the certification of randomness. What's more important, the random numbers are generated in the remote station without any assumptions about its device, observing the first steering-based random number extraction resisting the actual loss and noise. By reconstructing the covariance matrix (CM) of the shared entangled state with homodyne detection, the nonzero randomness is verified at the local station. The quantum random numbers with a generation rate of 7.06 Mbits/s are extracted from the measured amplitude quadrature fluctuation of the state owned by the remote party and pass the randomness statistical test suite. Our experiment presents an application of distributed CV EPR steering in fiber channel and makes a key step towards generating quantum random numbers for future network protocols[28,29].

**Results**

**Theoretical framework**

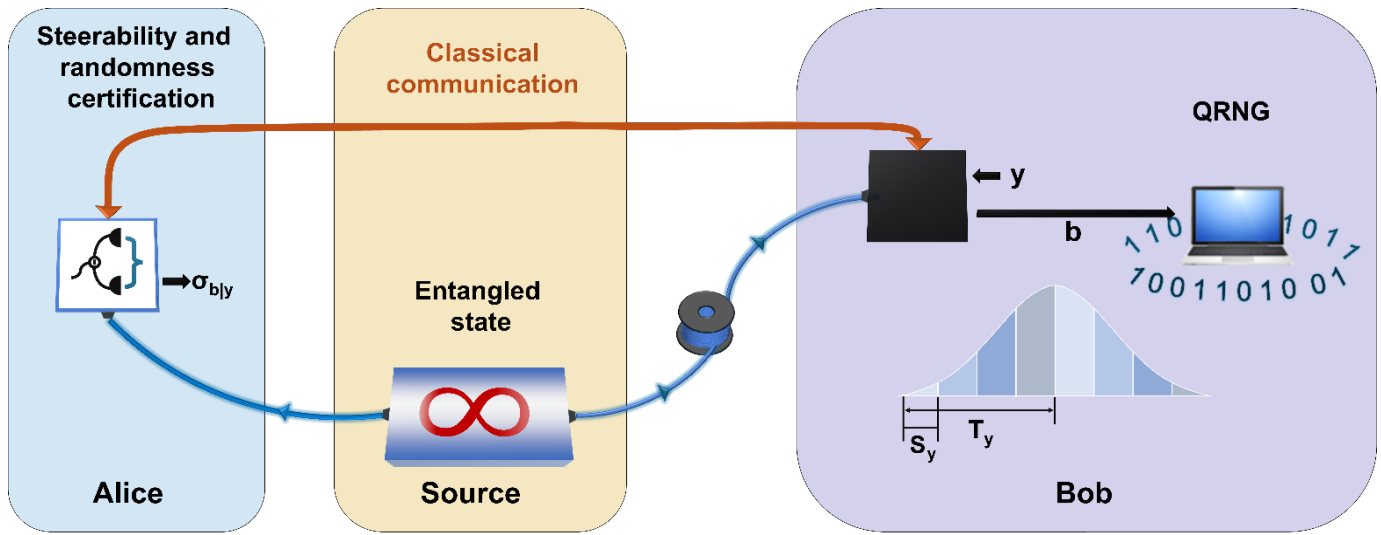

Fig. 1. Schematic of 1SDI QRNG. One mode of EPR entangled state is transmitted to Bob through a fiber channel. Alice's device is well-characterized (transparent box), while Bob's device is not (black box). Alice certifies the steerability and randomness based on her measurement results and the received Bob's results from the classical communication channel. Once the randomness is confirmed, Bob measures his own state again for QRNG.

As shown in Fig. 1, the 1SDI scenario we considered is close to the actual situation where only the device of local user, named Alice, is well-characterized (transparent box), while the device of the remote user, named Bob, is not (black box). One mode of a CV EPR entangled state is transmitted to Bob through a fiber (quantum) channel, while the other mode is kept by Alice. The whole protocol contains three steps: Step-I, EPR steering verification, where Alice and Bob verify the distributed steerability by measuring their own states; Step-II, randomness certification, where Alice verifies the steering-based randomness; Step-III, quantum random number extraction in Bob's station, by measuring his own state when Alice confirms the validity of steering-based randomness.

To be more specific, in the step-I, Alice and Bob perform a series of quadrature measurements on their own optical modes respectively and Bob sends his measurement results to Alice through the classical channel. Alice verifies the steerability based on the reconstructed CM from their measurement results (see details in Materials and methods).

The quadrature operators of each mode are denoted by $\hat{q}_i = \hat{a}_i^\dagger + \hat{a}_i$ and $\hat{p}_i = i(\hat{a}_i^\dagger - \hat{a}_i)$, where $\hat{a}_i^\dagger$ and $\hat{a}_i$ are the creation and annihilation operators and satisfy the canonical commutation relations $[\hat{a}_j, \hat{a}_k^\dagger] = 2\delta_{jk}$. Since the distributed entangled state and the measurements both have Gaussian nature, the entanglement properties of such system can be described by its CM with elements $\sigma_{ij} = \langle \hat{R}_i \hat{R}_j + \hat{R}_j \hat{R}_i \rangle / 2 - \langle \hat{R}_i \rangle \langle \hat{R}_j \rangle$, where the vector $\hat{R} = (\hat{q}_A, \hat{p}_A, \hat{q}_B, \hat{p}_B)^\top$ collects the quadrature operators for both modes. The steerability from Bob to Alice can be quantified by the parameter

$$\mathcal{G}^{B \to A} = \max\left\{0, \frac{1}{2}\ln\frac{\text{Det}\,\sigma_B}{\text{Det}\,\sigma_{AB}}\right\} \tag{1}$$

where $\sigma_B$ and $\sigma_{AB}$ denote the CM for the reduced mode $\hat{B}$, and the group $(\hat{A}\hat{B})$, respectively[30]. This quantity is a monotone under Gaussian local operations and classical communication and vanishes if and only if the state described by its covariance matrix is nonsteerable by Gaussian measurements. Once the steerability from Bob to Alice exists, the probability that Bob will obtain the results for his measurement cannot be explained by a local deterministic distribution, resulting in intrinsic randomness among his outcomes.

In the step-II, Alice estimates the amount of randomness based on the conditional states for Bob's specific measurements and outcomes. The outcomes of a homodyne detection, however, are continuous form over the whole phase space and cannot be directly used for randomness certification, so we bin them into a finite number of outcomes by utilizing the periodic coarse-grained protocol[31], as follows:

$$M_{b|y} = \int_{\mathbb{R}} f_b(z, T_y) |z\rangle_y \langle z| dz$$
$$f_b(z, T_y) = \begin{cases} 1, & bs_y \leq z \bmod T_y < (b+1)s_y \\ 0, & \text{otherwise} \end{cases} \tag{2}$$

Here the measurements $y$ performed by Bob are characterized as positive-operator-valued measures (POVMs) with finite elements $M_{b|y}$, where the outcomes $b \in \{0, 1, \cdots, o_B - 1\}$, $y \in \{\hat{q}, \hat{p}\}$ and $|z\rangle_y$ are the corresponding eigenstates. The periodic protocol defined by the periodic mask function is parameterized by the period $T_y$ with $o_B$ outcomes and bin width $s_y = T_y / o_B$, as illustrated in Fig. 1. After the strategy for binning Bob's results is determined, full information of the conditional states $\sigma_{b|y}^{\text{obs}}$ in Alice can be obtained from the reconstructed CM, as the CM contains all relevant information of the Gaussian state shared between them. The set of conditional states is called assemblage, and one can certify randomness associated with Bob's measurement outcomes through the analysis of it[32]. To be specific, the figure of merit we use to quantify the amount of randomness is the maximal probability $P_g(y^*)$ that an eavesdropper (Eve) correctly predicts the outputs when she knows the input $y^*$ and any other available side information, which can be formulated as a semidefinite program (SDP):

$$\max_{\sigma_{b|y}^e} \quad P_g(y^*) = \text{Tr}\left[\sum_e \sigma_{b=e|y^*}^e\right] \tag{3a}$$

$$\text{s.t.} \quad \sum_e \sigma_{b|y}^e = \sigma_{b|y}^{\text{obs}}, \quad \forall b, y \tag{3b}$$

$$\sum_b \sigma_{b|y}^e = \sum_b \sigma_{b|y'}^e, \quad \forall e, y \neq y' \tag{3c}$$

$$\sigma_{b|y}^{e} \geq 0, \quad \forall b, y, e \tag{3d}$$

Therefore, the amount of randomness generated by Bob's outcomes is quantified by the min-entropy[33] $H_{\min}(y^*) = -\log_2 P_g(y^*)$.

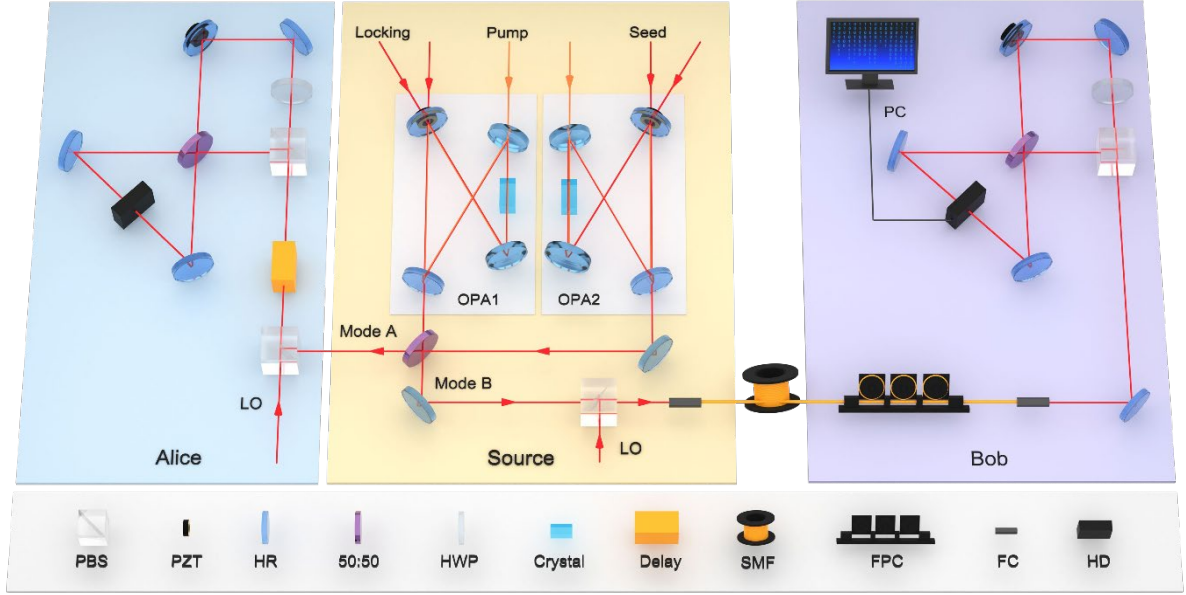

Fig. 2. Experimental setup for 1SDI QRNG through fiber channels. One mode of the EPR entangled state is distributed to Bob through a fiber channel and the other mode is retained and measured by Alice. After the certification of EPR steering and randomness at Alice's station, Bob extracts quantum random numbers independently and locally at his station. OPA: optical parametric amplifier, PBS: polarization beam-splitter, PZT: piezo-electric transducer, HR: high reflective mirror, 50:50: 50:50 beam-splitter, HWP: half-wave plate, SMF: single-mode fiber, FPC: fiber polarization controllers, FC: fiber coupler, HD: homodyne detector, LO: local oscillator, PC: personal computer.

Moreover, instead of full quantum state characterization, we can also certify randomness based on the joint probability distribution between Alice and Bob's limited measurements. The joint probability distribution $p^{\text{obs}}(ab|xy)$ can be derived from the reconstructed CM when only a few measurement directions at Alice's side are considered. The protocol of randomness certification based on the joint probability distribution is presented[27], by replacing Eq. (3b) with $\sum_e \text{Tr}\left[M_{a|x}\sigma_{b|y}^{e}\right] = p^{\text{obs}}(ab|xy), \quad \forall a,b,x,y$, for known elements $M_{a|x}$ of POVMs of Alice. It also guarantees that Eve's strategy is compatible with the observed probabilities.

In the step-III, Bob extracts quantum random numbers independently and locally at his station once Alice confirms that the steering-based randomness exists. Bob performs homodyne detection again to measure the fluctuations of amplitude or phase quadrature of his own optical mode and extract quantum random numbers from a large amount of measured fluctuations.

**The Experiment**

The schematic of the experimental setup of 1SDI QRNG is illustrated in Fig. 2. We generate the squeezed state of light at 1550 nm by using an optical parametric amplifier (OPA), which is a bow-tie traveling-wave cavity consisting of two plane mirrors, two concave mirrors, and a type-0 periodically poled potassium titanyl phosphate (PPKTP) crystal with a length of 10 mm. By controlling the relative phase difference between the pump and seed beams to zero, a quadrature phase squeezed state is obtained from the OPA. Two squeezed states with −2.78 dB/+3.47 dB and −2.69 dB/+3.47 dB squeezing/antisqueezing are prepared from OPA1 and OPA2, respectively. These squeezing levels of generated states are mainly limited by the normalized pump power and escape efficiency of the OPA, as well as the total detection efficiency. More details are given in Materials and methods. As purity of the squeezed state is more important for the amount of generated randomness than squeezing level, the pump power of OPAs is 20 mW which can achieve the highest squeezing purity in our experiment. The CV EPR entangled state is obtained by coupling these two phase squeezed states on a 50:50 beam splitter with a relative phase difference of $\pi/2$.

In the distribution of the EPR state, the polarization multiplexing technique is applied to transmit one mode of the EPR state and local oscillator (LO) through one fiber. The mode $\hat{B}$ of the CV EPR state with vertical polarization and a LO with horizontal polarization are coupled on a polarization beam-splitter (PBS) and the output is transmitted to Bob through a single-mode fiber (SMF) channel. After the transmission, the mode $\hat{B}$ and the LO are separated by a PBS at Bob's station to implement the homodyne detection.

The most challenging part in the distribution of EPR steering is the synchronization of two EPR entangled modes owned by Alice and Bob. Only the photons generated simultaneously present the highest quantum correlation, once the time delay between two photons exceeds the coherence time, the quantum correlation will decrease and even disappear. Thus, it is important to synchronize the photons in Alice's and Bob's station in the experiment. To do so, we transmit the EPR mode owned by Alice through the same fiber length (Delay in Fig. 2) to make sure the maximum quantum correlation is observed between Alice's and Bob's modes.

In the experiment of step-I, Alice and Bob measure the amplitude and phase quadratures of their own optical modes simultaneously and record the output of the homodyne detectors by a digital-storage oscilloscope with a sampling rate of 10 MS/s. Here, time-domain data is acquired by demodulating output signals of the homodyne detector at a sideband frequency of 4 MHz within a 300 kHz bandwidth and amplified 200 times via a low noise preamplifier. After the measurement, Bob sends his results of amplitude and phase quadrature of mode $\hat{B}$ to Alice through the classical channel. Alice reconstructs the CM based on both Bob's and her measurement results and verifies the steerability from Bob to Alice.

In the experiment of step-II, after the verification of EPR steering, Alice quantifies the amount of certifiable randomness with the optimized periodic coarse-grained protocol. As the state we generated is not pure, we choose the binning periods $T_{\hat{p}} = T_{\hat{q}}$ for Bob's two quadrature measurements $\{\hat{p},\hat{q}\}$, which provides higher randomness than the case satisfying mutual-unbiasedness condition[31], i.e., $T_{\hat{p}} = 2\pi o_B/T_{\hat{q}}$. Within the chosen binning periods, we bin their outputs into $o_B=32$ outcomes, corresponding to a fixed bits combination of length 5 assigned to each sample point[3]. Each outcome is associated with a conditional state $\sigma_{b|y}^{\text{obs}}$ at Alice's side, which can be obtained from the reconstructed CM. As we mentioned before, the condition (3b) can be replaced by the joint probabilities with Alice's limited measurements. We further bin the outputs of four measurement directions $\{\hat{p}_A,(\hat{p}_A+\hat{q}_A)/\sqrt{2},(\hat{p}_A-\hat{q}_A)/\sqrt{2},\hat{q}_A\}$. The output of each measurement is binned into $o_A=32$ outcomes, $o_A-1$ of which divides the range $[-5,5]$ evenly, and the last bin constitutes everything outside this range. For each coarse-grained protocol, one can describe the measurement $M_{a|x}$, $M_{b|y}$ performed by Alice and Bob. Based on the reconstructed CM, the conditional states $\sigma_{b|y}^{\text{obs}} = \text{Tr}_B[I \otimes M_{b|y}\rho]$ of Alice as well as the joint probability distribution $p^{\text{obs}}(ab|xy)$ are obtained. Hence the

randomness can be certified as described in Eq. (3).

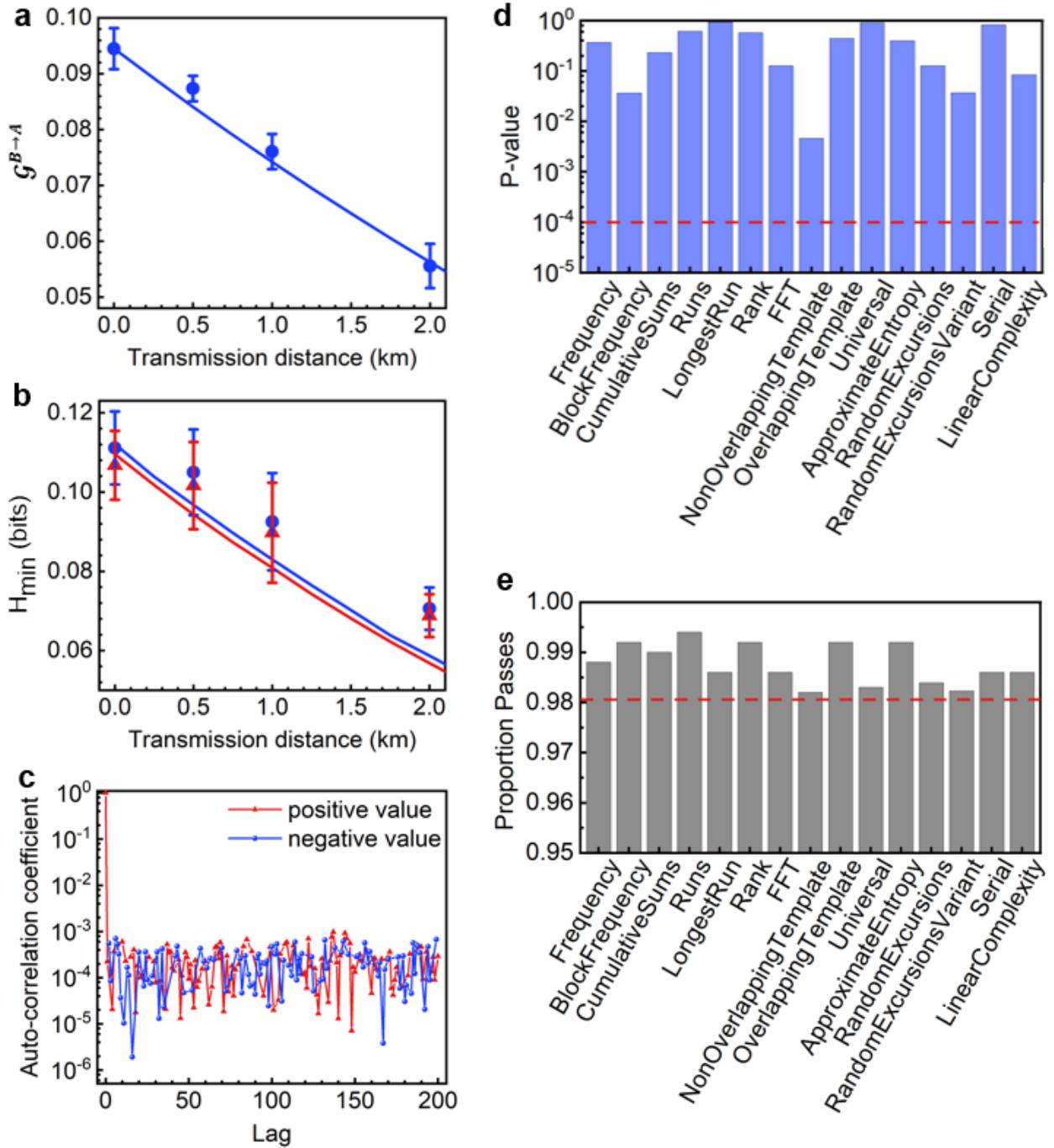

Fig. 3. Results of distributed quantum steering, certified randomness, and extracted quantum random numbers. **a** The measured steerability $\mathcal{G}^{B \to A}$ with transmission distance up to 2 km fiber. **b** The certified randomness $H_{min}$ at Alice's station. In all cases, the outputs of Bob's two quadrature measurements are binned into $O_B = 32$ outcomes. The min-entropy $H_{min}$ is quantified from the conditional states at Alice's side (blue curve and points) and joint probabilities of Alice's four measurements (red curve and points), respectively. Each experimental result point represents the mean value of three trails, and error bars correspond to one standard deviation from statistical data. The curve corresponds to the theoretical expectation. **c** Autocorrelation of the extracted random numbers with data size of 10 Mbits for the 2 km transmission. The absolute coefficients of positive and negative values are presented. **d** The P-value results from the 15 statistical tests conducted in the NIST test suite for the 2 km transmission distance. The extracted random numbers are divided into 1000 sequences with each sequence around 1 Mbits. Each test calculates 1000 p-values, and we use the chi-square test to calculate the final P-value. A sequence passes the test when its p-value is larger than 0.01. The obtained P-values are all larger than 0.0001, which confirms the uniformity of the extracted random numbers. **e** All the proportions of the sequences that pass the test beyond 0.9805607.

In the experiment of step-III, to extract quantum random numbers, Bob measures the amplitude quadrature of his own mode using a homodyne detector when Alice confirms the existence of randomness. Since the distribution of the measured amplitude fluctuation of the EPR mode is Gaussian rather than a uniform distribution, the raw random data cannot pass any randomness statistical tests[34]. To eliminate the classical effects and improve the statistical quality of quantum random numbers, we apply the Toeplitz hashing[34]. Here we acquire 1 Tbits raw data and extract 1 Gbits random numbers by Toeplitz matrix. Finally, these extracted random numbers are evaluated by using the standard randomness statistical test suite.

**Experimental results**

The result of EPR steering verification in the step-I is shown in Fig. 3a, where the Gaussian steerabilities $\mathcal{G}^{B \to A}$ for different transmission distances in the fiber channel are presented. The curves represent theoretical predictions based on the entangled state, which is generated by mixing two squeezed states originating from OPA1 and OPA2 on a balanced beam splitter. Additionally, the effect of fiber channel loss is also considered in the theoretical predictions by adopting the noisy channel described in the Materials and methods. It is obvious that the steerability decreases with the increase of transmission distance. The presence of the steerability from Bob to Alice guarantees the success of the whole protocol.

According to different coarse-grained protocols, we substitute the corresponding conditional states $\sigma_{b|y}^{\text{obs}}$ into Eq. (3) and obtain the randomness for different transmission distances (blue curve) in the step-II, as shown in Fig. 3b. As previously explained, we also compare the randomness with limited four measurement directions $\{\hat{p}_A, (\hat{p}_A + \hat{q}_A)/\sqrt{2}, (\hat{p}_A - \hat{q}_A)/\sqrt{2}, \hat{q}_A\}$ on Alice's side (red curve). The randomness derived from limited measurements is slightly lower than those achieved through conditional states. Nonetheless, the overall trend remains consistent, that is, the amount of certifiable randomness $H_{\min}$ decreases as the transmission distance increases.

In the extraction of random numbers at Bob's station, due to the difficulty of directly processing a long raw data sequence, we split the 1 Tbit raw random sequence into several $n$-bit-long blocks and process them individually[35]. For each block of raw data, we employ an $n \times m$ Toeplitz matrix as an extractor, where $n + m - 1$ pre-stored seeds true random bits are used to construct the Toeplitz matrix and m random bits are extracted by multiplying the Toeplitz matrix with $n$ raw bits. For a transmission distance of 2 km, the randomness $H_{\min}$ is certified to be 0.07057 bits per sample. Consequently, we choose $m = 1024$ and $n = 72,600 > m \times 5 / 0.07057 = 72,552$ to generate the corresponding extracted random bits[23,34], and obtain the quantum random numbers with a generation rate of 7.06 Mbits/s (see Table I for a summary of the QRNG demonstrations based on entangled states). Additionally, the generation rates of random numbers at 0.002, 0.5, and 1 km fiber channels are 11.11, 10.5, and 9.25 Mbits/s respectively (see details in the Supplementary information), which are slightly higher than that in a 2 km fiber channel, as the certifiable randomness $H_{\min}$ at these three transmission distances are larger.

After the extraction of the quantum random numbers, we analyze the autocorrelation coefficients on the postprocessed random numbers at a transmission distance of 2 km, as depicted in Fig. 3c. It can be found that the autocorrelation coefficients consistently approach zero at different time lags, with an average value of $1.28 \times 10^{-5}$. Finally, we test the extracted random numbers with the NIST test suite, which comprises 15 different tests. For these tests, we divide the extracted random numbers into 1000 sequences. The results of the NIST test suite for random bits at transmission distances of 2 km are presented in Figs. 3d and 3e, and the results at the other three transmission distances are given in Supplementary information. All the P-values exceed 0.0001, which demonstrates that the extracted random numbers satisfy a uniform distribution. The proportion passes of all sequences fall within the confidence interval of $0.99 \pm 0.0094392$, providing a strong evidence of the randomness of Bob's outcomes.

TABLE I. Generation rates of QRNG demonstrations based on entangled states. DI: Device independent, SDI: Source-device independent, 1SDI: One-sided device-independent. DoF: Degree of freedom used to encode information.

| QRNG | Type | DoF | Generation rate |
|---|---|---|---|
| 36 |  |  | 1.5 × 10$^{-5}$ bits/s |
| 37 |  |  | 0.4 bits/s |
| 11 | DI | DV | 114 bits/s |
| 38 |  |  | 13.5 kbits/s |
| 39 |  |  | 3.6 kbits/s |
| 22 | SDI | DV | 4 Mbits/s |
| This work | 1SDI | CV | 7.06 Mbits/s |

**Discussion**

In summary, we experimentally demonstrate the 1SDI QRNG based on the distributed deterministic CV quantum steering through a 2 km fiber channel. By reconstructing its covariance matrix, we verify the steerability from Bob to Alice at Alice's station first. Then the randomness is certified in the 1SDI scenario where the remote user lacks the ability to prepare quantum state and makes no additional assumptions about the device. Once the randomness exists, we extract the quantum random numbers locally at Bob's station by measuring the fluctuations of his own optical mode. The quantum random numbers with a generation rate of 7.06 Mbits/s are achieved at Bob's station which is 2 km away from the entanglement source.

The demonstrated protocol can be generalized to the quantum networks involving multiple users, where only a few users own trustworthy devices while the untrustworthy devices are used for the most users. The multipartite scenario can bring diverse structures and properties of the distributed steering, and hence additional protection against possible attacks can be found[40]. The transmission distance can be extended by further enhancing squeezing purity and reducing channel loss and excess noise. For instance, by enhancing squeezing purity, the transmission distance can be further improved to 11.42 km within the current experimental feasibility, since up to -10 dB/+11 dB of squeezing/antisqueezing has been demonstrated[41]. The presented results make a key step toward the application of EPR steering in the 1SDI scenario and have potential applications in fiber-based quantum networks.

**Materials and methods**

**Details of experiment**

The laser source employed in our experiment is a 1550 nm fiber laser (NKT Photonics). The OPA, which is in a bow-tie configuration, consists of a type 0 PPKTP crystal $(1 \times 2 \times 10 \text{ mm}^3)$, two plane mirrors, and two concave mirrors with $r = 50$ mm. Both the front and rear faces of the crystal are anti-reflectivity coated for both wavelengths to reduce the intra-cavity losses. A piezo-actuated plane mirror is used as the input mirror for the seed beam and locking beam, which is coated with high reflectivity (HR) at 1550 nm and 775 nm. The other plane mirror is used as the output mirror, which has a partial reflectivity of $87.5\% \pm 1\%$ for 1550 nm and HR for 775 nm. The pump beam is coupled into the OPA through a concave mirror, which is HR for 1550 nm and has a $80\%$ reflectivity for 775 nm. The other concave mirror is HR for both wavelengths. The cavity length and linewidth of the OPA are about 478 mm and 12.3 MHz, respectively. For performing the measurements, we carefully adjust the OPA to the desired state through mode matching, temperature control, gain adjustment, and phase locking to ensure the stability of the squeezing source.

To lock the cavity of the OPA, a counter-propagating locking beam carrying on a phase modulation signal of 16.67 MHz is injected into the OPA through the input mirror with PZT. By detecting the transmitted locking beam with a photodetector and feed-backing the demodulated error signal to the PZT on the input mirror, the cavity length is stabilized. The relative phase difference between the seed and pump beams of the OPA is actively stabilized to $0°$ by feed-backing the error signal, which is obtained from the interference signal of the seed and down-converted beams to a PZT-mounted mirror in the path of the seed beam. After locking the cavity and relative phase difference, a phase-squeezed state is deterministically prepared. For the homodyne detector, the seed beam power is 100 μW in the OPA and the LO power is 600 μW. The variances of the squeezing and antisqueezing quadratures are measured with a total detection efficiency of $87\%$. Other related parameters for the measurement are given in the Supplementary Information.

The transmission efficiency of the fiber channel, denoted as $\eta$, is defined by $\eta = \eta_0 \times 10^{-\alpha L/10}$, where $\eta_0 = 0.9$ represents the fiber coupling efficiency, $L$ is the length of fiber transmission, and $\alpha$ is the loss coefficient of the SMF, typically ~ 0.2 dB/km at 1550 nm. Here the modes transmitted through the noisy channel are

$$\hat{q}_i^{out} = \sqrt{\eta}\hat{q}_i^{in} + \sqrt{1-\eta}(\hat{q}_N + \hat{q}_v)$$
$$\hat{p}_i^{out} = \sqrt{\eta}\hat{p}_i^{in} + \sqrt{1-\eta}(\hat{p}_N + \hat{p}_v) \quad (4)$$

where the subscripts $N$ and $v$ represent the noise and vacuum state respectively. The excess noise in the fiber channel is given by $\delta = \Delta^2\hat{q}_N = \Delta^2\hat{p}_N$. The power of the local oscillator distributed in the fiber channel is set to 0.6 mW, which leads to the excess noise of $\delta = 0.01$ shot-noise-unit in our experiment.

In the process of extraction, to obtain the seed sequence for the Toeplitz hashing extractor, we discretize the probability distribution of the measured amplitude fluctuations of mode $\hat{B}$ into the binary raw data first. Then we choose 50 Mbits data and feed them to the SHA-512 function. In this way, we obtain the seed sequence of 0.22 Mbits.

**Reconstruction of the covariance matrix**

The CM of the EPR entangled state can be partially characterized as (assuming the cross-correlations between different quadratures are zero):

$$\sigma = \begin{bmatrix} \Delta^2\hat{q}_A & 0 & C(\hat{q}_A,\hat{q}_B) & 0 \\ 0 & \Delta^2\hat{p}_A & 0 & C(\hat{p}_A,\hat{p}_B) \\ C(\hat{q}_A,\hat{q}_B) & 0 & \Delta^2\hat{q}_B & 0 \\ 0 & C(\hat{p}_A,\hat{p}_B) & 0 & \Delta^2\hat{p}_B \end{bmatrix}$$

where $\Delta^2\hat{q}_A$, $\Delta^2\hat{p}_A$, $\Delta^2\hat{q}_B$, $\Delta^2\hat{p}_B$ represent the variances of amplitude and phase quadratures of EPR entangled modes $\hat{A}$ and $\hat{B}$, respectively, which are obtained from the measured $\hat{q}_{A(B)}$ and $\hat{p}_{A(B)}$ in time domain. $C(\hat{q}_A,\hat{q}_B)$ and $C(\hat{p}_A,\hat{p}_B)$ represent cross-correlations between the output optical modes and are calculated based on the measured variances with the following relations[42]:

$$C(\hat{R}_i,\hat{R}_j) = \frac{1}{2}\left[\Delta^2(\hat{R}_i + \hat{R}_j) - \Delta^2\hat{R}_i - \Delta^2\hat{R}_j\right]$$
$$C(\hat{R}_i,\hat{R}_j) = -\frac{1}{2}\left[\Delta^2(\hat{R}_i - \hat{R}_j) - \Delta^2\hat{R}_i - \Delta^2\hat{R}_j\right]$$

When the transmission distances between the source and Bob are 0.002, 0.5, 1, and 2 km, the reconstructed CMs of the EPR entangled state are:

$$\begin{bmatrix} 1.35 \pm 0.01 & 0 & 0.77 \pm 0.01 & 0 \\ 0 & 1.36 \pm 0.01 & 0 & -0.79 \pm 0.01 \\ 0.77 \pm 0.01 & 0 & 1.38 \pm 0.01 & 0 \\ 0 & -0.79 \pm 0.01 & 0 & 1.33 \pm 0.01 \end{bmatrix}$$

$$\begin{bmatrix} 1.32 \pm 0.01 & 0 & 0.73 \pm 0.02 & 0 \\ 0 & 1.35 \pm 0.01 & 0 & -0.76 \pm 0.03 \\ 0.73 \pm 0.02 & 0 & 1.35 \pm 0.00 & 0 \\ 0 & -0.76 \pm 0.03 & 0 & 1.33 \pm 0.01 \end{bmatrix}$$

$$\begin{bmatrix} 1.32 \pm 0.00 & 0 & 0.73 \pm 0.01 & 0 \\ 0 & 1.34 \pm 0.01 & 0 & -0.74 \pm 0.01 \\ 0.73 \pm 0.01 & 0 & 1.35 \pm 0.00 & 0 \\ 0 & -0.74 \pm 0.01 & 0 & 1.32 \pm 0.01 \end{bmatrix}$$

$$\begin{bmatrix} 1.30 \pm 0.01 & 0 & 0.69 \pm 0.01 & 0 \\ 0 & 1.31 \pm 0.00 & 0 & -0.68 \pm 0.01 \\ 0.69 \pm 0.01 & 0 & 1.33 \pm 0.01 & 0 \\ 0 & -0.68 \pm 0.01 & 0 & 1.30 \pm 0.01 \end{bmatrix}$$

respectively.

## Acknowledgements

This work is supported by the National Natural Science Foundation of China (Grants No. 11834010, No. 11975026, No. 12125402, and No. 12350006), and the Innovation Program for Quantum Science and Technology (Grant No. 2021ZD0301500), the Beijing Natural Science Foundation (Grant No. Z240007), the Fundamental Research Program of Shanxi Province (Grant No. 20210302121002), and the Fund for Shanxi "1331 Project" Key Subjects Construction.

## Author contributions

X. Su, Q. He and Y. Xiang conceived the original idea; J. Zhang, M. Zhao and X. Su designed the experiment and carried out the experiment; Y. Li and Y. Xiang completed the theoretical analysis; J. Zhang and Y. Li analyzed the data; D. Han, J. Liu, and M. Wang participated in part of the experiment. Q. He, Q. Gong and X. Su managed the project. J. Zhang, Y. Li, Y. Xiang, X. Su, and Q. He prepared the paper.

## Data availability

The data that support the findings of this study are available from the corresponding author upon reasonable request.

## Conflict of interest

The authors declare no competing interests.

∗ These authors contributed equally

† xiangy.phy@pku.edu.cn

‡ suxl@sxu.edu.cn

## Reference

1. Metropolis, N. & Ulam, S. The Monte Carlo method. *Journal of the American Statistical Association* **44**, 335-341 (1949).
2. Herrero-Collantes, M. & Garcia-Escartin, J. C. Quantum random number generators. *Reviews of Modern Physics* **89**, 015004 (2017).
3. Gabriel, C. et al. A generator for unique quantum random numbers based on vacuum states. *Nature Photonics* **4**, 711-715 (2010).
4. Born, M. Zur quantenmechanik der stoßvorgänge. *Zeitschrift für Physik* **37**, 863-867 (1926).
5. Jennewein, T. et al. A fast and compact quantum random number generator. *Review of Scientific Instruments* **71**, 1675-1680

(2000).

6. Lin, X. et al. Security analysis and improvement of source independent quantum random number generators with imperfect devices. *npj Quantum Information* **6**, 100 (2020).
7. Bell, J. S. On the Einstein Podolsky Rosen paradox. *Physics Physique Fizika* **1**, 195-200 (1964).
8. Einstein, A., Podolsky, B. & Rosen, N. Can quantum-mechanical description of physical reality be considered complete? *Physical Review* **47**, 777-780 (1935).
9. Bierhorst, P. et al. Experimentally generated randomness certified by the impossibility of superluminal signals. *Nature* **556**, 223-226 (2018).
10. Liu, Y. et al. Device-independent quantum random-number generation. *Nature* **562**, 548-551 (2018).
11. Liu, Y. et al. High-speed device-independent quantum random number generation without a detection loophole. *Physical Review Letters* **120**, 010503 (2018).
12. Shen, L. J. et al. Randomness extraction from Bell violation with continuous parametric down-conversion. *Physical Review Letters* **121**, 150402 (2018).
13. Zhang, Y. B. et al. Experimental low-latency device-independent quantum randomness. *Physical Review Letters* **124**, 010505 (2020).
14. Horodecki, R. et al. Quantum entanglement. *Reviews of Modern Physics* **81**, 865-942 (2009).
15. Brunner, N. et al. Bell nonlocality. *Reviews of Modern Physics* **86**, 419-478 (2014).
16. Wang, M. et al. Asymmetric quantum network based on multipartite Einstein–Podolsky–Rosen steering. *Journal of the Optical Society of America B* **32**, A20-A26 (2015).
17. Guo, Y. et al. Experimental measurement-device-independent quantum steering and randomness generation beyond qubits. *Physical Review Letters* **123**, 170402 (2019).
18. Wang, J. W. et al. Multidimensional quantum entanglement with large-scale integrated optics. *Science* **360**, 285-291 (2018).
19. Joch, D. J. et al. Certified random-number generation from quantum steering. *Physical Review A* **106**, L050401 (2022).
20. Xu, F. H., Shapiro, J. H. & Wong, F. N. C. Experimental fast quantum random number generation using high-dimensional entanglement with entropy monitoring. *Optica* **3**, 1266-1269 (2016).
21. Máttar, A. et al. Experimental multipartite entanglement and randomness certification of the W state in the quantum steering scenario. *Quantum Science and Technology* **2**, 015011 (2017).
22. Zhang, J. N. et al. Realization of a source-device independent quantum random number generator secured by nonlocal dispersion cancellation. *Advanced Photonics* **5**, 036003 (2023).
23. Zhang, Q. et al. Quantum random number generator based on twin beams. *Optics Letters* **42**, 895-898 (2017).
24. Marangon, D. G., Vallone, G. & Villoresi, P. Source-device-independent ultrafast quantum random number generation. *Physical Review Letters* **118**, 060503 (2017).
25. Michel, T. et al. Real-time source-independent quantum random-number generator with squeezed states. *Physical Review Applied* **12**, 034017 (2019).
26. Avesani, M. et al. Source-device-independent heterodyne-based quantum random number generator at 17 Gbps. *Nature Communications* **9**, 5365 (2018).
27. Ioannou, M. et al. Steering-based randomness certification with squeezed states and homodyne measurements. *Physical Review A* **106**, 042414 (2022).
28. Kimble, H. J. The quantum internet. *Nature* **453**, 1023-1030 (2008).
29. Wehner, S., Elkouss, D. & Hanson, R. Quantum internet: a vision for the road ahead. *Science* **362**, eaam9288 (2018).
30. Kogias, I. et al. Quantification of Gaussian quantum steering. *Physical Review Letters* **114**, 060403 (2015).
31. Tasca, D. S. et al. Mutual unbiasedness in coarse-grained continuous variables. *Physical Review Letters* **120**, 040403 (2018).
32. Passaro, E. et al. Optimal randomness certification in the quantum steering and prepare-and-measure scenarios. *New Journal of Physics* **17**, 113010 (2015).
33. Brown, P. J., Ragy, S. & Colbeck, R. A framework for quantum-secure device-independent randomness expansion. *IEEE Transactions on Information Theory* **66**, 2964-2987 (2020).
34. Ma, X. F. et al. Postprocessing for quantum random-number generators: entropy evaluation and randomness extraction. *Physical Review A* **87**, 062327 (2013).

35. Zhao, Z. H., Ma, X. F. & Zhou, H. Y. Performance optimization on practical quantum random number generators: modification on min-entropy evaluation and acceleration on post processing. Print at https://doi.org/10.48550/arXiv.2011.04130 (2020).
36. Pironio, S. et al. Random numbers certified by Bell's theorem. *Nature* **464**, 1021-1024 (2010).
37. Christensen, B. G. et al. Detection-loophole-free test of quantum nonlocality, and applications. *Physical Review Letters* **111**, 130406 (2013).
38. Liu, W. Z. et al. Device-independent randomness expansion against quantum side information. *Nature Physics* **17**, 448-451 (2021).
39. Shalm, L. K. et al. Device-independent randomness expansion with entangled photons. *Nature Physics* **17**, 452-456 (2021).
40. Li, Y. et al. Randomness certification from multipartite quantum steering for arbitrary dimensional systems. *Physical Review Letters* **132**, 080201 (2024).
41. Vahlbruch, H. et al. Detection of 15 db squeezed states of light and their application for the absolute calibration of photoelectric quantum efficiency. *Physical Review Letters* **117**, 110801 (2016).
42. Steinlechner, S. et al. Strong Einstein-Podolsky-Rosen steering with unconditional entangled states. *Physical Review A* **87**, 022104 (2013).